\documentstyle{lamuphys}

\begin{document}

\font\grand=cmbx10 at 12pt 
\font\eightrm=cmr8
\font\eightbf=cmbx8
\font\eightit=cmti8
\font\eightsl=cmsl8
\font\eighteuf=eufm8
\font\tensmc=cmcsc10
\let\sc=\tensmc
\def\eightpoint{\let\rm=\eightrm \let\bf=\eightbf \let\it=\eightit
\let\sl=\eightsl \baselineskip = 9.5pt minus .75pt  \rm}
\def \smalltype{\let\rm=\eightrm  \let\bf=\eightbf
\let\it=\eightit \let\sl=\eightsl \let\mus=\eightmus
\baselineskip = 9.5pt minus .75pt  \rm}
\font\teneuf=eufm10  \font\seveneuf=eufm7 \font\fiveeuf=eufm5
\newfam\euffam \def\gr{\fam\euffam\teneuf}
\def\frak{\fam\euffam\teneuf}
\textfont\euffam=\teneuf \scriptfont\euffam=\seveneuf 
\scriptscriptfont\euffam=\fiveeuf
\def \scst {\scriptstyle}
\def \dst {\displaystyle}
\def \res {{\rm res}}
\def \newline{\hfill \break}
\def \oc {{\otimes }}
\def \di{\partial}
\def \Lg {\widetilde{\gr g}}
\def \LG {\widetilde{\gr G}}
\def \LGp {\widetilde{\gr G}_+}
\def \LGm {\widetilde{\gr G}_-}
\def \Lgp {\widetilde{\gr g}_+}
\def \Lgm {\widetilde{\gr g}_-}
\def \LgA {\widetilde{\gr g}_A}
\def \lgp {\widetilde{\gr g}_+}
\def \lgm {\widetilde{\gr g}_-}
\def \gl {{\gr gl}}  
\def \gA {{\bf g}_{A}}
\def \gB {{\bf g}_{B}}
\def \gBA {{\bf g}_{A}^B}
\def \gAB {{\bf g}_{B}^A}
\def \a {\alpha}
\def \b {\beta}
\def \d {\delta}
\def \g {\gamma}
\def \k {\kappa}
\def \eps {\epsilon}
\def \l {\lambda}
\def \o {\omega}
\def \m {\mu}
\def \s  {\sigma}
\def \th  {\theta}
\def \G {\Gamma}
\def \tr{{\rm tr}}
\def \Stab{{\rm Stab}}
\def \det{{\rm det}}
\def \Ad{{\rm Ad}}
\def \log {{\rm log}}
\def \red {{\rm red}}
\def \diag{{\rm diag}}
\def \ra {\rightarrow}
\def \lra{\longrightarrow}
\def \mt {\mapsto}
\def \lmt {\longmapsto}
\def \wt {\widetilde}
\def \ss{\subset}
\def \k {\kappa}
\def \d {\delta}
\def \a {\alpha}
\def \t {\tau}
\def \eps {\epsilon}
\def \IBA {{\cal I}_A^B}
\def \IAB {{\cal I}_B^A}
\def \AA {{\cal A}}
\def \DD {{\cal D}}
\def \HH {{\cal H}}
\def \NN {{\cal N}}
\def \MM {{\cal M}}
\def \OO {{\cal O}}
\def \PP {{\cal P}}
\def \bfI {{\bf I}}
\def \bff {{\bf f}}
\def \bfg {{\bf g}}
\def \bfv {{\bf v}}
\def \bfF {{\bf F}}
\def \bfG {{\bf G}}
\def \bfK {{\bf K}}
\def \scst {\scriptstyle}
\def \dst {\displaystyle} 
%
\def\Proclaim#1:#2\par{\smallbreak\noindent{\sc #1:\ }
{\sl #2}\par\smallbreak}
\def\Demo#1:#2\par{\smallbreak\noindent{\sl #1:\ }
{\rm #2}\par\smallbreak}
%
\title{Hamiltonian Dynamics, Classical $R$--matrices
           and Isomonodromic Deformations }
\titlerunning{Classical $R$--matrices and Isomonodromic Deformations}
\vskip -1 true in
\author{ J.~Harnad\inst{1,2}}
\institute{Department of Mathematics and Statistics, Concordia
University, 7141 Sherbrooke W., Montr\'eal, Canada H4B 1R6,  \and 
 Centre de recherches math\'ematiques, Universit\'e de Montr\'eal 
  C.~P. 6128-A, Montr\'eal, Canada H3C 3J7}
\maketitle
\vskip -6.5cm
\rightline{CRM-2511 (1997) 
\footnote{\eightpoint Text of invited talk presented at the workshop {\it
Supersymmetry and Integrable Models} held at the University of Illinois,
Chicago Circle, June\. 12--14, 1997. To appear in: Springer Lecture Notes
in Physics (1997/98).} 
\break}
\rightline{solv-int/9710012 \break} \bigskip
\vskip 6.5cm
\begin{abstract}
The Hamiltonian approach to the theory of dual isomonodromic deformations
is developed within the framework of  rational classical $R$--matrix 
structures on loop algebras. Particular solutions to the isomonodromic
deformation equations appearing in the computation of correlation functions in
integrable quantum field theory models are constructed through the
Riemann-Hilbert problem method. The corresponding
$\tau$--functions are shown to be given by the Fredholm determinant of a
special class of integral operators.
\end{abstract}
\section{Monodromy Preserving Hamiltonian Systems}
\renewcommand{\theequation}{1.\arabic{equation}}\setcounter{equation}{0}
\subsection{Isomonodromic Deformation Equations}

Monodromy preserving deformations of  rational covariant
derivative  operators of the form:
\begin{equation}
{\cal D}_{\l} := {\di \over \di \l} - {\cal N}(\l), \label{(1.1)}
\end{equation}
 where
\begin{equation}
{\cal N}(\l) := B + \sum_{i=1}^n{N_i \over \l- \a_i},  \label{(1.2)}
\end{equation} 
$B$ is the diagonal $r\times r$ matrix
\begin{equation}
B=\diag(\b_1, \dots, \b_r), \label{(1.3)}
\end{equation}
and the matrices $\{N_i\}_{i=1\cdots n}$
are $r\times r$ matrix functions of  the $n+r$ deformation parameters $\{a_i,
\b_a\}$, were studied by Jimbo {\it et. al.} in (\cite{[JMMS]}), 
(\cite{[JMU]}), (\cite{[JM]}).
It was shown there that the most general differentable monodromy
preserving deformations of such operators are determined by the integrable
Pfaffian system:
\begin{equation}
dN_i = -\sum_{j=1 \atop j \ne i}^n [N_i, N_j] d\log(\a_i - \a_j) 
    -[N_i, d(\a_i B) + \Theta],  \label{(1.4)}
\end{equation}
where $\Theta$ is the $r\times r$ matrix with elements
\begin{equation}
\Theta_{ab} =(1 -\d_{ab})(\sum_{i=1}^nN_i)_{ab} d\log(\b_a -\b_b). 
\label{(1.5)}
\end{equation}

  Such operators and their monodromy are of great importance in the theory of
quantum integrable systems, since the computation of correlation functions in
such systems very often leads to particular solutions to such systems 
(\cite{[IIKS]}), (\cite{[KBI]}). 
In the following subsection, it will be shown how these equations
may be understood as a compatible set of nonautonomous Hamiltonian
systems generated  by commuting Hamiltonians that are spectral invariants of
the matrix $\NN(\l)$. In subsequent sections, the classical
$R$--matrix approach to such systems will be explained and the computation of
certain solutions related to Fredholm determinant calculations via the matrix
Riemann--Hilbert problem will be decribed.
\subsection{Nonautonomous Hamiltonian Structure}
 
We begin with the Lie Poisson structure on  $(\oplus_{i=1}^n \gl(r))^*$, 
defined by  the following Poisson brackets between the various matrix elements
\begin{equation}
\{(N_i)_{ab},(N_j)_{cd}\} =\d_{ij}\left( \d_{bc}(N_i)_{ad} -
 \d_{ad} (N_i)_{cb}\right). \label{(1.6)}
\end{equation}
The system (\ref{(1.4)}) can be expressed in multi--Hamiltonian form by 
introducing the following Hamiltonian $1$--form on the parameter space
\begin{equation}
\theta := \sum_{i=1}^n H_id\a_i + \sum_{a=1}^r K_a d\b_a   \label{(1.7)}
\end{equation}
where
\begin{eqnarray}
H_i &:=& \tr(B N_i) + \sum_{j=1 \atop j\ne i}^n {\tr(N_iN_j) \over \a_i -\a_j}
  ,\qquad i=1, \dots, n \label{(1.8a)} \\
 K_a &:=& \sum_{i=1}^n \a_i (N_i)_{aa} + 
\sum_{b=1 \atop b \ne a}^r {(\sum_{i=1}^n N_i)_{ab}
 (\sum_{j=1}^n N_j)_{ba}  \over \b_a -\b_b}, \qquad a =1, \dots, r. 
\label{(1.8b)} 
\end{eqnarray}
Equations (\ref{(1.4)}) may then equivalently be written in multi--Hamiltonian 
form as
\begin{equation}
dN_i=\{N_i, \theta\}. \label{(1.9)}
\end{equation}
Involutiveness of the $H_i$'s and $K_a$'s, which will be explained in the 
following section, then implies that the differential form $\theta$ on the
parameter space is in fact closed
\begin{equation}
d\theta =0, \label{(1.10)}
\end{equation}
which allows one  to introduce the $\t$--function (cf.~\cite{[JMMS]}, 
\cite{[JMU]}) by the 
formula
\begin{equation}
\theta = d\log \tau. \label{(1.11)}
\end{equation}

  In the next section, we show how the above Hamiltonian structure may very 
naturally be viewed as the restriction of the rational $R$--matrix structure on
a loop algebra to a finite dimensional Poisson  submanifold. From this 
the commutativity  of the Hamiltonians $(H_i, K_a)$ defined above follows.
\section{Loop Algebra Moment Maps, Spectral Invariants and 
Isomonodromic Deformation Equations} 
\renewcommand{\theequation}{2.\arabic{equation}}\setcounter{equation}{0}
  The following discussion is based on the approach developed in 
\cite{[H]}.

\subsection{Dual Moment Maps and Split $R$-matrix Structure}

We introduce an auxiliary symplectic vector space $(M, \o)$, which will be 
referred to as the {\it generalized Moser space}, consisting of pairs $(F,G)$
of complex $N\times r$  matrices whose elements are viewed as canonically
conjugate variables. Thus, the symplectic form $\o$ is just
\begin{equation}
\omega := \tr(dF \wedge dG^T).  \label{(2.1)}
\end{equation}
We use the following notation to denote the loop algebra of $r\times r$ 
matrices depending on a loop parameter $\l$, viewed as a point on a circle
$S^1$ in the  complex
$\l$--plane, and its splitting into negative a positive Fourier components
\begin{eqnarray}
\wt{\gl}(r) = \wt{\gl}(r)_{+} + \wt{\gl}(r)_{-} &\sim& \wt{\gl}(r)^* 
\label{(2.2a)}\\
\wt{\gl}(r)_{\pm}^* &\sim& \wt{\gl}(r)_{\mp} \label{(2.2b)}
\end{eqnarray}%
The identification with the dual space indicated in eq.~(\ref{(2.2a)})
is defined through the Ad--invariant scalar product
\begin{equation}  
<X_1, X_2> = \oint_{S^1} \tr\left(X_1(\l)X_2(\l)\right)d\l.   \label{(2.3)}
\end{equation} 

  The rational $R$--matrix structure is obtained by just redefining the Lie 
algebra structure in such a way that the new algebra splits into a Lie
algebraic direct  sum of the positive and negative frequency parts, with a
change of sign in the Lie product for the second summand
\begin{equation}
\wt{\gl}_R(r) := \wt{\gl}(r)_{+} \ominus \wt{\gl}(r)_{-} \sim \wt{\gl}_R(r)^* . 
\label{(2.4)}
\end{equation}
The rational $R$--matrix structure is then just the  corresponding Lie Poisson
structure on $\wt{\gl}_R(r)^*$. Expressed in terms  of  the individual matrix
elements this gives
\begin{equation}
\left[\NN_{ab}(\l),\ \NN_{cd}(\mu)\right]={\d_{ad}(\NN_{cb}(\l) - 
\NN_{cb}(\mu))
-(ad \leftrightarrow cb)\over \l -\mu}. \label{(2.5)}
\end{equation}
This can be expressed more succintly in the tensorial (St. Petersburg)
notation as follows
\begin{equation}
\left\{\NN(\l) \oc \NN(\mu)\right\}= 
\left[r(\l-\mu), \ \NN(\l)\otimes {\bf I} +{\bf I}\otimes \NN(\mu)\right],
\label{(2.6)}
\end{equation}
where $\NN(\l)$ and $\NN(\mu)$ are viewed as endomorphisms of ${\bf C}^n $ and 
the rational $R$--matrix $r(\l -\mu)$ is the endomorphism of ${\bf C}^n \otimes
{\bf C}^n$ defined by
\begin{equation}
r(\l -\mu):= {P_{12} \over \l-\mu} \in {\rm End }({\bf C}^n \otimes {\bf C}^n),
\label{(2.7)}
\end{equation}
where $P_{12}$ denotes the endomorphism that interchanges the first and second
factors in ${\bf C}^n \otimes {\bf C}^n$.

   Now let $A$ and $B$ be the diagonal $N\times N$ and $r\times r$ matrices,
respectively, defined by
\begin{equation}
A :=\diag (\a_i) \in \gl(N),  \qquad
B :=\diag(\b_a) \in \gl(r),  \label{(2.8)}
\end{equation}
where the eigenvalues $\{\a_i\}_{i=1\cdots n}$ have multiplicities 
$\{k_i\}_{i=1, \dots n}$, and the eigenvalues $\{\b_a\}_{a=1\cdots r}$ are
multiplicity free. Define the Poisson subspace $\gAB \ss \wt{\gl}_R(r)^*$ by
\begin{equation}
\gAB:=\{\NN(\l) = B + \sum_{i=1}^n {N_i \over \l -\a_i}, \ N_i \in \gl(r)\} 
\sim
\sum_{i=1}^n \gl^*(r). \label{(2.9)}
\end{equation}
Then the following defines a Poisson quotient map of the symplectic space $M$, 
such that the image is identified with a Poisson submanifold of $\gAB$.
\begin{eqnarray}
\wt{J}^A_B: M &\lra&  \gAB   \nonumber \\
\wt{J}^A_B :(F,G) &\lmt& B+ G^T(A-\l I_r)^{-1}F  =:\NN(\l) \label{(2.10a)}\\
\NN (\l) &=& B+\sum_{i=1}^n {N_i \over \l - \a_i} \in
\wt{\gl}^*(r)_R   \label{(2.10b)} \\
 N_i &:=& -G^T_i F_i , \qquad F_i, G_i \in M^{k_i \times r}. \label{(2.10c)}
\end{eqnarray}%

   We may now apply the standard classical $R$--matrix theory to deduce a set 
of commuting Hamiltonian flows on the space $M$, generated by the spectral
invariant Hamiltonians on $\wt{\gl}(r)^*$, pulled back to $M$ via the above
Poisson map, for which the Hamiltonian flows are represented by Lax equations.
The resulting Hamiltonian flows are therefore isospectral for the matrix
$\NN(\l)$.  We denote by 
\begin{equation}
\IAB := {\cal I}(\wt{\gl}(r)^*)\vert_{\gAB}    \label{(2.11)}
\end{equation}
the ring of spectral invariants restricted to $\gAB$.
Then the classical $R$--matrix theory in this case tells us that:
\begin{enumerate}
\item[(i)] $\IAB$ is Poisson commutative.
\item[(ii)] For $H \in \IAB$, Hamilton's equations have the Lax form:
\end{enumerate}
\begin{eqnarray}%
 {d{\cal N} \over dt} &=& [\AA^H_\s,\ {\NN}], \label{2.12a}\\ 
\AA^H_\s &:=& \s dH_+(\NN) + (\s-1)dH_-(\NN) =: \PP_\s(dH(\NN)),  
 \label{2.12b}
\end{eqnarray}%
where the subscripts $\pm$ denote projections to the positive and negative 
Fourier components and $\s \in {\bf R}$ is arbitrary. 

  \noindent It follows that the spectral curve defined by the characteristic
equation
\begin{equation}
\det (\NN(\l)  -z I_r) =0,  \label{(2.13)}
\end{equation}
where 
\begin{equation}
\NN(\l)  B + G^T(A-\l I_N)^{-1}, \label{(2.14)}
\end{equation}
is invariant under the resulting Hamiltonian flows.

   In order to apply this to the isomonodromic deformation equations considered
above, we must adapt these results to the case of nonautonomous Hamiltonian
systems, in which the  flow parameters are reinterpreted as deformation
parameters upon which the spectral invariant Hamiltonians may explicitly
depend. This will be done in the next  subsection.

\subsection{Nonautonomous systems: isomonodromic deformations}

Letting the matrices $A$ and $B$ depend explicitly on some deformation
parameter $t$
\begin{equation}
A=A(t), \quad B=B(t),  \label{(2.15)}
\end{equation}
the above Lax equations must be modified to take the resulting explicit
$t$--dependence of the matrix $\NN(\l)$ into account.
This gives the nonautonomous system
\begin{equation}
{d{\cal N} \over dt} = [A^H_\s, {\cal N}] + {\di\NN \over \di t}.
\label{(2.16)}
\end{equation}
Suppose now that, for some $\s$ and $H$, the following special condition holds:
\begin{equation}
{\di\NN \over \di t} = {\di\AA^H_\s \over \di \l}.  \label{(2.17)}
\end{equation}
It follows that Hamilton's equations become {\it isomonodromic 
deformation} equations, since eq.~(\ref{(2.16)}) then takes the form of
commutativity conditions
\begin{equation}
\left[\DD_\l, \ \DD_t\right] =0 ,    \label{(2.18)}
\end{equation}
where
\begin{eqnarray}%
\DD_\l &:=& {\di \over \di \l} - \NN(\l) \label{(2.19a)}\\
\DD_t &:=& {\di \over \di t} - \AA^H_\s. \label{(2.19b)} 
\end{eqnarray}%
These are precisely the necessary conditions for the invariance of the
monodromy of the operator $\DD_\l$ under deformations in the parameter $t$.

   To apply this to the system (\ref{(1.4)}), we choose the following set 
of spectral invariant Hamiltonians $\{H_i \in \IAB\}_{i=1, \dots n}$
\begin{equation}
H_i({\cal N}) := {1\over 4 \pi i}\oint_{\l =\a_i}\tr(({\cal N}(\l))^2 d\l  
=\tr(BN_i) +\sum_{j=1 \atop j\ne i}^n{\tr(N_iN_j)\over \a_i - \a_j}.
\label{(2.20)}
\end{equation}
The autonomous form of Hamilton's equations that result are then
\begin{equation}
{\di{\cal N} \over \di t_i} = - [(dH_i)_-, {\cal N}],  \label{(2.21)}
\end{equation} 
where
\begin{equation}
(dH_i)_- = {N_i \over \l - \a_i} \in \wt{\gl}(r)_-. \label{(2.22)}
\end{equation}
Identifying the various deformation parameters now with the eigenvalues of the 
matrix $A$, $\{t_i =\a_i\}_{i=1,\dots, n}$, the Lax equations are modified
to the following form:
\begin{equation} 
{\di {\cal N} \over \di \a_i} = -[(dH_i)_-,{\cal N}]
 -{\di (dH_i)_- \over \di\l\quad}.   \label{(2.23)}
\end{equation} 
These are just the commutativity conditions
\begin{eqnarray}
[{\cal D}_\l, {\cal D}_i]& =&0, \quad  i=1, \dots, n, \label{(2.24a)}\\
{\cal D}_{\l} &:=& {\di \over \di \l} - {\cal N}(\l) \label{(2.24b)}\\
{\cal D}_i &:=& {\di \over \di \a_i} + (dH_i)_- ={\di \over \di \a_i}
 + {N_i \over \l - \a_i}. \label{(2.24c)}
\end{eqnarray}
guaranteeing the preservation of the monodromy of the operator $\DD_\l$ under 
the deformations generated by varying the $\a_i$'s. 
Evaluating residues at $\{\a_i\}_{i=1, \dots n}$ gives
\begin{eqnarray}
{\di N_j\over \di \a_i}&=& {[N_j,N_i]\over \a_j - \a_i}, \quad j \neq i,
 \quad i,j = 1, \dots, n, \label{(2.25a)} \\
{\di N_i\over \di \a_i} &=& [B +\sum_{j=1 \atop j\neq i}^n{N_j \over 
\a_i - \a_j}, N_i].  \label{(2.25b)} 
\end{eqnarray}
which are just the $\a_i$ components of the differential system (\ref{(1.4)}).

\subsection{Dual Isomonodromic System}

  To obtain the $\b_a$ components of this system, it is convenient to 
introduce another representation, in terms of a second system of rational
covariant  derivative operators whose monodromy will also be preserved: the
{\it dual} isomonodromic system. Define another loop algebra $\wt{\gl}(N)$,
consisting of $N\times N$ matrices depending  similarly on a loop parameter $z$
that lies on a circle $S^1$ in the complex $z$--plane, with corresponding
splitting into positive and negative Fourier components: 
\begin{equation}
\wt{\gl}(N) =\wt{\gl}(N)_+  +\wt{\gl}(N)_- \sim \wt{\gl}(N)^*,  \label{(2.26)}
\end{equation}
and corresponding rational $R$--matrix structure $\wt{\gl}(N)_R^*$ on the 
dual space. We also define a corresponding Poisson subspace
$\gBA \ss \wt{\gl}(N)_R^*$  consisting of rational elements
$\MM(z)$ of the form
\begin{equation}
\gBA := \{{\cal M}(z) = -A + \sum_{a=1}^r {M_a\over{z-\b_a}}\} 
\sim \sum_{a=1}^r \gl^*(N).  \label{(2.27)}
\end{equation}
Introduce the ``dual'' Poisson map from $M$ to $\gBA$ as:
\begin{eqnarray}
\wt{J}^B_A : M &\lra& \gBA  \nonumber\\
\wt{J}^B_A : (F,G) &\lmt& -A - F(B -z I_N)^{-1}G^T := {\cal M}(z),
\label{(2.28)}
\end{eqnarray}
and denote the dual ring of spectral invariants restricted to $\gBA$ as
\begin{equation}
\IBA := {\cal I}(\wt{\gl}(N)^*)\vert_{\gBA}.   \label{(2.29)}
\end{equation}
We then have the remarkable fact that the spectral rings $\IBA$ and $\IAB$ {\it
coincide} when pulled back under the respective Poisson maps $\wt{J}^{B}_A$ 
and
$\wt{J}^{A}_B$ to $M$. 
\Proclaim Theorem 2.1 (duality theorem): The two spectral invariant rings
$\wt{J}^{B*}_A(\IBA)$ and 
$\wt{J}_B^{A*}(\IAB)$ {\it coincide.}
 
\Demo Proof: This essentially follows from the simple linear algebra identity
\begin{eqnarray}
&&\det(A-\l I_N) \det ( B + G^T(A-\l I_N)^{-1}F  -z I_r) \nonumber\\
&= &
\det(B - zI_r) \det (A + F(B-zI_r)^{-1}G^T - \l I_N),
\label{(2.30)}
\end{eqnarray}
which implies that the spectral curves of $\NN(\l)$ and $\MM(z)$ are identical.

    Now define the set of spectral invariant Hamiltonians $\{ K_a \in
\IBA\}_{a=1, \dots r}$, similarly to the $H_i$'s, as:
\begin{equation}
K_a := {1\over 4 \pi i} \oint_{z=\b_a} \tr ({\cal M}(z))^2dz .  \label{(2.31)}
\end{equation}  
On $\gBA$, these similarly generate the equations:
\begin{equation}
{\di {\cal M} \over \di \b_a} = -[(dK_a)_-,{\cal M}]
 -{\di (dK_a)_- \over \di z },   \label{(2.32)}
\end{equation}
where
\begin{equation}
(dK_a)_-(z) =  {M_a\over z-\b_a} \in \wt{\gl}(N)_-,     \label{(2.33)}
\end{equation}
which imply the invariance of the monodromy of the rational covariant 
derivative operator
\begin{equation}
\DD_z := {\di \over \di z} - \MM(z)   \label{(2.34)}
\end{equation}
under the deformations generated by changes in the parameters 
$\{\b_a\}_{a=1, \cdots r}$.

 Using Theorem 2.1, we may also pull back the $K_a$'s under the map
$\wt{J}^{A}_B$, to determine  corresponding spectral invariant Hamiltonian 
functions  of $\NN(\l)$. These again generate isomonodromic deformation
equations for the operator $\DD_\l$, which may be expressed
\begin{equation}
{\di {\cal N}\over \di \b_a} = \left[ (dK_a)_+, {\cal N}\right]
+{\di (dK_a)_+ \over \di \l\quad} ,  \label{(2.35)}
\end{equation}
where
\begin{equation}
(dK_a)_+(\l) = \l E_a + \sum_{b=1 \atop b\ne a}^r {\sum_{i=1}^n }
{E_a N_i E_b + E_bN_i E_a \over \b_a -\b_b},   \label{(2.36)}
\end{equation}
and $E_a$ denotes the elementary $r\times r$ matrix with diagonal entry $1$ in 
the $aa$ position and zero elsewhere. Evaluating residues at $\l =\a_i$ gives
\begin{equation}
{\di N_i\over \di \b_a} = \left[ (dK_a)_+(\a_i), N_i\right],   \label{(2.37)}
\end{equation}
which are precisely the $\b_a$ components of the isomonodromic
system  (\ref{(1.4)}). Similarly, viewing $\{H_i\}_{i=1, \dots n}$ as
Hamiltonians defined on $\gBA$, these  generate the {\it dual} equations, 
which imply the invariance of the monodromy of the  operator $\DD_z$:
 \begin{equation}
{\di {\cal M}\over \di \a_i} = \left[ (dH_i)_+, {\cal M}\right]
+{\di (dH_i)_+ \over \di z\quad},   \label{(2.38)}
\end{equation}
where
\begin{equation}
(dH_i)_+(z) = -z E_i + \sum_{j=1 \atop j\ne i}^n {\sum_{a=1}^r }
{E_i M_a E_j + E_jM_a E_i \over \a_i - \a_j},   \label{(2.39)}
\end{equation}
and $E_i$ denotes the elementary $n\times n$ matrix with entry $1$ in the 
$ii$ position.
 Evaluating residues at  $z=\b_a$ gives these in the form
\begin{equation}
{\di M_a\over \di \a_i} = \left[ (dH_i)_+(\b_a), M_a\right] .  \label{(2.40)}
\end{equation}

\section{Isomonodromic Deformations and the 
Riemann~--Hilbert Problem} 
\renewcommand{\theequation}{3.\arabic{equation}}\setcounter{equation}{0}
In this last section, we discuss certain specific solutions of the 
above isomonodromic deformation equations which can be constructed through
application of the Zakharov--Shabat ``dressing'' method (\cite{[NZMP]}). This
class of solutions is of particular interest from the viewpoint of
applications, since they arise in the  calculation of correlation functions for
quantum integrable systems (\cite{[KBI]}), (\cite{[HI]}) and spectral
distributions in the theory of  random matrices (\cite{[TW]}), (\cite{[HTW]}).
The results quoted in this  section are based on joint work
with A. Its; the full details may be found in the joint paper 
(\cite{[HI]}).

  The particular class of solutions to the isomonodromic deformation equations
in question may be constructed by applying the {\it dressing method}, based on
the matrix Riemann--Hilbert problem, suitably adapted to this case. To do this,
we first introduce a {\it vacuum} solution $\Psi_0$, which is chosen as the
invertible $r\times r$ matrix function obtained by exponentiating
 $\l B$
\begin{equation}
\Psi_0(\l) := e^{\l B}.   \label{(3.1)}
\end{equation}
We then introduce a family of loop group elements $H_0(\l)$, viewed as 
$Gl(r)$--valued functions defined along some oriented, closed curve $\G$
chosen, in this case, to  pass {\it through} the points $\{\a_i\}$,
consecutively, with the latter ordered by their subscript labels. We also
assume in the following that the number $n$ of such points is even, and write
$n=2m$. (If the number happens to be odd, we just increase it by adding
$\l=\infty$ as the last point.) Let $\{\G_j\}_{j=1\cdots m}$ denote the segment
of
$\G$ between $\a_{2j-1}$ and $\a_{2j}$ and let $\th_j(\l)$ denote the
characteristic function, along $\G$, of the interval $\G_j$. We define
$H_0(\l)$ as the piecewise constant element of the form
\begin{equation}
H_0(\l) := \bfI_r +2\pi i\sum_{j=1}^n \bff_j \bfg_j^T\th_j(\l),  \label{(3.2)}
\end{equation}
where $\{\bff_j, \bfg_j\}_{j=1 \cdots m}$  is any fixed set of
$r \times p$ complex, rectangular matrices, with $p\le r$, satisfying the null
conditions
\begin{equation}
\bfg_j^T\bff_k=0, \quad \forall j, k.  \label{(3.3)}
\end{equation}
 The relevant matrix Riemann--Hilbert problem consists of finding a nonsingular
$r\times r$ matrix valued function $\chi (\l )$ that is analytic on the 
complement of $\G$, extending to $\l=\infty$ off $\G$, with asymptotic form  
\begin{equation}
\chi(\l) \sim \bfI_r + \OO\left(\l^{-1}\right)  \label{(3.4)}
\end{equation}
for $\l \ra \infty$, and has cut discontinuities across $\G$ given by
\begin{equation}
\chi_-(\l) = \chi_+(\l) H(\l), \quad \l \in \G,  \label{(3.5)}
\end{equation}
where $\chi_+(\l)$ and $\chi_-(\l)$ are the limiting values of $\chi(\l)$ as 
$\G$ is approached from the left and the right, respectively, and $ H(\l)$ is
the $r\times r$ invertible matrix valued function along $\G$ defined by
\begin{equation}
H(\l) = \Psi_0(\l) H_0(\l) \Psi_0^{-1}(\l).  \label{(3.6)}
\end{equation}
Following the Zakharov--Shabat dressing method, we define the {\it dressed}
wave function as
\begin{equation}
\Psi_{\pm}(\l):= \chi_{\pm}(\l)\Psi_0(\l),  \label{(3.7)}
\end{equation}
with limiting values $\Psi_{\pm}$ on either side of the segments of $\G$ given
by
\begin{equation}
\Psi_{\pm}(\l):= \chi_{\pm}(\l)\Psi_0(\l).  \label{(3.8)}
\end{equation}
We then have the following result, which is quoted here from \cite{[HI]},
\Proclaim Theorem 3.1: The wave function $\Psi(\l)$ defined by (\ref{(3.8)})
satisfies the equations
\begin{eqnarray}
{\di \Psi \over \di \l} - 
\left(B  + \sum_{j=1}^n {N_j \over \l - \a_j}\right)\Psi &=&0, 
\label{(3.9a)}\\
{\di \Psi \over \di \a_j}+ {N_j \over \l -\a_j}\Psi&=&0, \label{(3.9b)}
\end{eqnarray}
with the $N_j$'s given by
\begin{equation}
N_j := -\bfF_j \bfG^T_j,  \label{(3.10)}
\end{equation}
where
\begin{equation}
\bfF_j:= \lim_{\l \ra \a_j} \bfF(\l) \quad
\bfG_j:= (-1)^{j}\lim_{\l \ra \a_j} \bfG(\l).  \label{(3.11)}
\end{equation}
 This implies the commutativity
\begin{equation}
[\DD_\l, \ \DD_{\a_j}]=0,\quad   [\DD_{\a_i}, \ \DD_{\a_j}]=0, 
\quad i, j=1, \dots, n  \label{(3.12)}
\end{equation}
of the operators
\begin{eqnarray}
\DD_\l &:=& {\di \over \di \l} - B - \sum_{j=1}^n {N_j \over \l -\a_j}
={\di \over \di \l} - \NN(\l) \label{(3.13a)}\\ 
\DD_{\a_j} &:=& {\di \over \di \a_j} + {N_j \over \l -\a_j},\label{(3.13b)}
\end{eqnarray}
and hence the invariance of the monodromy data of the operator $\DD_\l$  under
changes in the parameters $\{\a_j\}$ . 

Thus, the operators defined in eqs.~(\ref{(3.13a)})-(\ref{(3.13b)}) 
represent a solution to the
$\a_j$ components of the isomonodromic system (\ref{(1.4)}). 
The following result,
also quoted from \cite{[HI]}, shows that the same construction also provides a
solution to the $\b_a$ components.
\Proclaim  Theorem 3.2:  The wave function $\Psi(\l)$ also satisfies the
equations 
\begin{equation}
\DD_{\b_a} \Psi = 0, \quad a=1, \dots, r, \label{(3.14)}
\end{equation}
where the operators $\{\DD_{\b_a}\}_{b=1, \dots, r}$ are defined by
\begin{equation}
\DD_{\b_a}:= {\di \over \di \b_{a}} - \l E_{a} -  \sum_{b=1 \atop b\neq a}^r
 {E_a \left(\sum_{j=1}^nN_j\right) E_b 
+ E_b \left(\sum_{j=1}^nN_{j} \right)E_a \over \b_a - \b_b},   \label{(3.15)}
\end{equation}
with the $N_i$'s given by eqs.~(\ref{(3.10)}), (\ref{(3.11)}). 
This implies the commutativity conditions
\begin{equation}
\left[\DD_\l, \ \DD_{\b_a}\right] =0,  \quad 
\left[\DD_{\b_a}, \ \DD_{\b_b}\right] =0,\quad a,b=1, \dots, r, 
\label{(3.16)}
\end{equation}
and hence the invariance of the monodromy data of $\DD_\l$  under the
deformations  parameterized by $\{\b_a\}_{a=1, \dots, r}$.

  From the viewpoint of applications to quantum integrable systems 
(\cite{[IIKS]}, \cite{[KBI]}) and the spectral theory of random matrices 
(\cite{[TW]}, \cite{[HTW]}) this
construction has particular importance, since the underlying $\t$--function,
as defined in eq.~(\ref{(1.11)}), is just the Fredholm determinant of an integral
operator that may be constructed from the same data, and which gives the
correlation functions and spectral distribution generating functions in
question. This result is contained in the following theorem, also quoted
from \cite{[HI]}.
\Proclaim Theorem 3.3:
Let $\bfK$ be the $p \times p$ matrix Fredholm integral operator acting on 
${\bf C}^p$--valued functions ${\bf v}(\l)$,
\begin{equation}
\bfK ({\bfv})(\l) = \int_\G K(\l, \mu) {\bfv}(\mu)d\mu, \label{(3.17)}
\end{equation}
defined along the curve $\G$, with integral kernel given by
\begin{equation}
K(\l, \mu) = {\bff^T(\l)\bfg(\mu)\over \l - \mu},  \label{(3.18)}
\end{equation}
where $\bff, \bfg$ are the rectangular $r\times p$ matrix valued functions
\begin{eqnarray}
\bff(\l)&:=&   \Psi_0(\l)\sum_{j=1}^m \bff_j \th_j(\l) \label{(3.19a)}\\
\bfg(\l)&:=& \left(\Psi_0^T(\l)\right)^{-1}\sum_{j=1}^m \bfg_j \th_j(\l).
\label{(3.19b)} 
\end{eqnarray}
Then the logarithmic derivative of the Fredholm determinant is given by
\begin{equation}
d\ln \det (\bfI - \bfK)  = \o =\sum_{k=1}^n H_k d\a_k + \sum_{a=1}^r K_ad\b_a,
\label{(3.20)}
\end{equation}
where the individual factors may be expressed
\begin{eqnarray}
H_k &=& {\di \ln \det (\bfI - \bfK) \over \di \a_{k}} 
= \tr(BN_{k}) + \sum_{j=1, j\neq k}^{n}{\tr{N_{j}N_{k}}\over
 {\a_{k}-\a_{j}}} \label{(3.21a)}\\ 
K_a & =& {\di \ln \det (\bfI - \bfK) \over \di \b_{a}} \nonumber\\
&=&  \sum_{j=1}^{n}\a_{j}\left(N_{j}\right)_{aa} 
+ \sum_{b=1 \atop b\neq a}^r
{ {\left (\sum_{j=1}^{n}N_{j}\right )_{ab}\left (\sum_{k=1}^{n}N_{k}\right
)_{ba}} \over{\b_{a}-\b_{b}}}.
\label{(3.21b)}
\end{eqnarray}
Hence, $\det (\bfI - \bfK)$ may be identified as the $\t$--function defined
in eq.~(\ref{(1.11)}). 

  Finally, it should be mentioned that the dual isomonodromic systems defined
in eqs.~(\ref{(2.34)})--(\ref{(2.40)}) may be derived in exactly the same way,
by interchanging the r\^oles of the matrices $A$ and $B$ when defining the 
vacuum wave function (\ref{(3.1)}). The corresponding curve $\wt{\G}$ must  be
chosen, in the complex $z$--plane, so as to pass through the  parameters
$\{\b_a\}$ giving the diagonal elements of the matrix $B$. The resulting
$\t$--function turns out to just be given by the Fredholm determinant of the
{\it dual} Fredholm integral operator $\wt{\bfK}$, which is related to the
operator $\bfK$ appearing in Theorem 3.2 by taking a Fourier--Laplace transform
along the curves $\G$ and $\wt{\G}$. Full details regarding this result, as
well as a number of related results, including generalizations to isomonodromic
deformations of operators having higher order pole singularities, may be found
in \cite{[HI]}.
\bigskip

\noindent
{\Large \bf  Acknowledgements.} 
This research was supported in part by the Natural Sciences and
Engineering Research Council of Canada and the Fonds FCAR du Qu\'ebec.


%
\end{document}